# Efficient Power Flow Management and Peak Shaving in a Microgrid-PV System

Sakshi Mishra
Transmission Planning Engineer,
American Electric Power
Tulsa, USA
sakshi.m@outlook.com

Praveen Palanisamy
Alumni, Robotics Institute,
Carnegie Mellon University
Pittsburgh, USA
praveen.palanisamy@outlook.com

*Abstract*— With the increasing penetration of the roof-top solar PV and the rising interest in net-zero energy homes concept, there is a need of balancing the performance of intelligent controllers, their cost-effectiveness and over-all sophistication of the microgrid systems in order to manage the bi-directional power flow in the small as well as large size microgrids. This paper proposes solutions to efficiently manage power flow and to achieve peak shaving in a renewable-source fed microgrid system. The paper details the design and simulation of a photovoltaic source fed microgrid system that achieves peak shaving and efficient power flow management using advanced metering and a smart control unit. The proposed system enables microgrid to maintain the power consumption within limits during peak hours by shedding luxurious loads automatically. Under the grid-connected mode of the microgrid, the system feeds the excess power available to the utility grid during lower load requirements and withdraws the power deficit from the grid during high demand hours when photovoltaic power generation is not sufficient to fulfill the load requirement. The proposed system exhibits desirable power flow management performance, and is capable of functioning with distributed generation sources other than PV.

## I. INTRODUCTION

Microgrid systems play an important role in boosting the reliability of electricity supply by utilizing on-site renewable generation [1]. Traditionally, supervisory control and data acquisition (SCADA) architectures [2] have been used to manage energy resources, but these centralized architectures may have limited functionalities in their application to micro-grid. With the proliferation of renewable energy sources and their increased penetration in existing grid, the power flow has become bidirectional and complexity of power system has increased manifolds. A study of the International Energy Agency concludes that, although Distributed Generation (DG) has higher capital costs than power plants, it has potential and that it is possible with DG to supply all the demand with same reliability, but with lower capacity margins [3]. A Smart Grid is a means of delivering power to the consumers with increased reliability, greater flexibility and yet reasonable costs. The grid is thus evolving into smart grid with the proliferation of sensor (resulting in increased visibility into grid operation), specialized controllers and faster communication pathways. The main advantages of a smart grid are enunciated in [4], which includes infrastructure with zero or less carbon emissions, improved quality of energy supply etc. The concept of smart grid and microgrid are intertwined when it comes to operating a microgrid in its grid-connected mode, which calls for a smart-control architecture, which can manage the bi-directional flow of power between the smart grid and the microgrid. The smart control of the microgrid is proposed in [5] to have a cyber-layer and a physical layer and cyber layer is further categorized as control layer and application layer of smart control [6]. Smart micro grids generate, distribute, and regulate the flow of electricity to consumers locally. Peak load is one of the major issues for Independent System Operators (ISO), so is with islanded operation of microgrid. A microgrid is capable of supplying a rated amount of power. If a situation occurs where the peak load demand exceeds the power supply capacity, then it leads to the breakdown of the microgrid system. Thus, power distribution companies often have to procure additional power from the market at a very high cost to meet the demand during peak hours. Consequently, consumers are charged at an increased tariff during peak times. Batteries [7] and emergency generators [8] have been used to fulfill the demand during peak hours. However, Omni-presence of battery storage to manage all supply-demand disturbances is still a pipedream. This paper presents a grid-synchronized smart microgrid-PV system, which uses distributed grid intelligence (DGI) to enable easy monitoring of the micro-grid along with a Smart Control Unit (SCU). The SCU enables the system to achieve peak shaving in a smarter way under islanded mode and control the power flow for efficient energy management during grid-connected mode. The load shedding method is beneficial to both the energy producers and consumers- it enables the producer to stay within the contract's maximum demand and can optimize the network. For the consumers, it reduces the electricity bills by reducing the power consumption during the peak tariffs. Improvement of the overall utilization of the system resources and increase in the capacity factor of the transmission line and distribution facilities can be achieved by reduction in peak load. Thus, the proposed system converts the micro-grid into a smart grid which promises widespread load response, especially during peak load conditions and a reliable and efficient energy management. The system proposes a control mechanism in which, the luxurious load are disconnected automatically from the microgrid during peak load hours in the islanded mode of operation, with the help of the SCU and AMI resulting in peak shaving. Hence, the proposed system eliminates the need of costly and uneconomical peak power plants and provides a means to save energy.

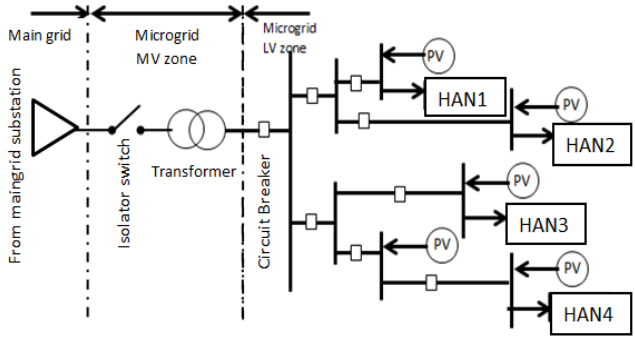

**Figure 1 Block Diagram of proposed Microgrid Architecture**

The proposed system is implemented using MATLAB/SIMULINK software package. The rest of the paper is organized as follows. Section 2 gives an overview of the model of the proposed system with grid-synchronized microgrid, which is having photo voltaic (PV) as the source of generation. Section 3 delineates the functioning of the smart control unit, which enables system to achieve peak shaving. Section 4 discusses the peak shaving concept. Section 5 details the modeling of the micro-grid system with its smart features. Section 6 manifests the simulations and results. Section 7 describes the key benefits of the proposed system followed by conclusion in Section 8.

## II. OVERVIEW OF THE SYSTEM

The block diagram of the grid-synchronized microgrid is shown in Fig.1. It has five photovoltaic systems as the distributed generation sources, each rated at 20KW giving a combined power rating of 100KW. The proposed system presents a novel way of peak shaving and a smart power flow management in two different modes of operation of microgrid in order to increase the energy and cost efficiency of the microgrid:

### A. Grid connected mode

In grid connected mode, when the PV power generation is in excess of the demand, it supplies the loads and the remaining power is fed into the utility grid. If the PV power generation is less than the demand, then the additional required power is drawn from grid.

### B. Islanded mode

In islanded mode, if the load demand exceeds the generation capacity of the renewable energy source then peak shaving is implemented.

## III. SMART CONTROL UNIT

Smart Control Unit (SCU) is the heart of the system, which proposes an approach to decentralize the control. Further, with bi-directional communication between the utility grid and the consumer, this decentralized control becomes a smart control. The SCU is represented in Fig. 2. The need for automation has increased with the dynamic nature of load variation in the smart grid and SCU fulfills this need efficiently. The two essential parts of the proposed Smart Control system are Advanced Metering Infrastructure and Head End System (HES).

### A. Advanced Metering Infrastructre

AMI is an integration [9] of technologies, which provides an intelligent connection between consumers and system operators. It supports automated metering of the energy consumed, transfers data from meter to HES, enables the consumer to keep track of the energy consumption and cost incurred in real time, and provides time stamped system information. The two main components in AMI are smart meter and communication pathway.

### B. Smart Meter

It measures the continuous energy consumption of a house. The measured data is displayed in the meter enabling the consumer to keep track of their power consumption. Thus, the consumer no longer stays as a "passive" consumer rather they become "active" consumers who can influence the pricing level by manipulating their demands. Based on single-phase and 3-phase connection, the single-phase and three-phase smart meter will measure the total power consumption of a house. A 3-phase energy meter having high linearity is required for accurate power measurement. A highly integrated, multi-channel, multi-bit, analog front-end 3-phase energy meter having isolation amplifier for preventing magnetic interference [10] is required for this purpose. The block diagram representation of the metering infrastructure is shown in Fig.2. Single phase Smart meters are also employed to measure active energy, reactive energy, apparent energy, active power, RMS voltage, RMS current, power factor, temperature and frequency [11]. Also smart energy meters are installed at generation sites where renewable sources of energy (PV in the proposed system) are used, as the generation is dependent on environmental conditions.

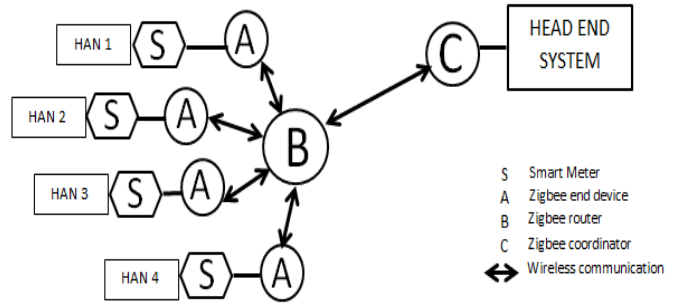

**Figure 2 Smart Control Unit**

- *Communication Pathway*: The data measured by the Smart meter will be transmitted wirelessly from the consumer end to HES. The generated power will also be sent to HES. The technology which is preferred for wireless communication is ZigBee, which is low power wirelessly networking standard designed for controlling and monitoring actions.

- *Head End System*: The HES acquires the data using the communication pathway and then constantly monitors the fluctuations of the generation and consumption of power. It calculates the difference between load requirement and generation of the PV array and depending upon their difference it either draws from or feeds power to the utility grid. HES have the control of domestic load in HAN which enables it to have the smart functionality of Peak Shaving also. Thus Smart Control Unit is used to achieve the goals of peak shaving and power flow management by forming the link between the cyber layer and physical layer of the microgrid.

## IV. PEAK SHAVING

Peak shaving is appropriate when the total load on a site is exceeding the agreed maximum demand, which otherwise penalizes the consumers to pay extra charges during peak hours. If this problem is unattended, it may lead to drop in frequency, voltage fluctuation and damage to system equipment. Frequency being a system wide variable affects the overall operation of the system. Generally, peak shaving can be done either by using peak power plants or by load shedding.

### A. Use of Peaking Plants

During peak time, these plants start to operate to fulfill the increased demand. As peak load plants run for a short or highly variable time, it is not economical for producers to establish efficient peak load power plants. Thus, the power supplied by these plants is costlier. Also the plants which are used as peak load plants like diesel plants and gas turbines burn natural gas which causes environmental pollution because of high carbon emissions.

### B. Load Shedding

The proposed mechanism of peak shaving economizes production cost and electricity bills. Hence, it is considerably advantageous for power producers and consumers. In the current scenario, when there is a peak demand or overload on the system and no peak power plant is available to fulfill the demand then the whole system is shut down leading to a blackout. The proposed system introduces a smart way to shed some luxurious loads automatically at critical loading values in such a way that the system is not overloaded and microgrid can operate without the risk of breakdown under islanded mode.

## V. MODELING OF THE SYSTEM

The grid connected microgrid and HAN system is modeled using MATLAB/SIMULINK. The modeling of the different components in the proposed system is described in this section.

### A. Photo Voltaic (PV) System

Each PV array as in Fig. 1 is designed for 20 KW maximum power output. An incremental conductance method is employed to get the maximum power from the PV arrays. Boost converter is employed to regulate and boost the output voltage of the PV arrays. A three phase three-level Voltage Source Inverter (VSI) converts the DC power into three phase AC. The components of the PV system are described in the following sections.

#### a) PV array

A two diode model is used to represent a PV cell which has better accuracy at low irradiance level [12]. The diode current is modeled using (1)

$$I = I_{ph} - I_{s1}\left[e^{\frac{V+IR_S}{V_t}} - 1\right] - I_{s2}\left[e^{\frac{V+IR_S}{AV_t}}\right] - \frac{V+IR_S}{R_p} \quad (1)$$

where, $V_t = \frac{kT}{e}$; $I_{ph}$ - photo-generated current; $I_{s_1}$ - diode saturation current of first diode; $I_{s_2}$ - diode saturation current of second diode; A - diode parameter; $R_s$ - series resistance; $R_p$ - parallel resistance.

$I_{ph}$, $I_{s_1}$, $I_{s_2}$, A, $R_s$ and $R_p$ are dependent on irradiance and temperature. Each 20 KW PV-array consists of 13 strings connected in parallel. Each string has 5 series connected modules rated at 305.2 W and one module has 96 cells. Module specification under STC is $[V_{oc}, I_{sc}, V_{mp}, I_{mp}]$ = [64.2, 5.96, 54.7, 5.58].

#### b) MPPT Controller

The MPPT controller takes PV voltage $V_{pv}$ and PV $I_{pv}$ current as inputs and tracks the maximum power point and adjusts the duty cycle of the boost converter in such a way that maximum power is extracted from PV array. Variable step size Incremental Conductance (IC) algorithm [13] is used for MPPT. In this method, the controller measures the incremental changes in array current and voltage, to predict the effect of voltage change. IC algorithm is used because it can continue tracking maximum power with high efficiency during fast irradiance changes and also it determines the Maximum Power Point without oscillating around the maximum value unlike in other methods.

#### c) Boost converter

A switched inductor boost converter (SIBC) is employed [14]. In SIBC, the inductor of the boost converter is replaced with a switched inductor branch. Consequently, the conversion gain ratio as in (2) can be increased. With its duty cycle adjusted by MPPT controller, boost converter takes $V_{pv}$ and regulates the output voltage of PV array. The gain of SIBC as in (2), is more than the gain of traditional boost converter by a factor of (1+D). The small signal model of boost converter is given by (3) and (4).

$$\frac{V_o}{V_{in}} = \frac{1+D}{1-D}$$

(2)

$$\frac{\delta\begin{bmatrix}i_l(t)\\v_o(t)\end{bmatrix}}{\delta t} = [A]\begin{bmatrix}i_l(t)\\v_o(t)\end{bmatrix} + [B]\begin{bmatrix}v_g(t)\\d_1(t)\end{bmatrix}$$

(3)

Where,

$$A = \begin{bmatrix} 0 & -\left(\frac{1-D}{2*L}\right) \\ \left(\frac{1-D}{C}\right) & -\left(\frac{1}{RC}\right) \end{bmatrix}; \quad B = \begin{bmatrix} \left(\frac{1+D}{2*L}\right) & \left[\frac{v_o}{2*L} + \frac{v_g}{2*L}\right] \\ 0 & -\left(\frac{i_l}{C}\right) \end{bmatrix}$$

(4)

Where, $i_l$ - inductor current; $v_o$ - output voltage of the boost converter; $v_g$ - input voltage to the boost converter; D- duty cycle; C- capacitance; L-inductance; R-resistance.

#### d) Voltage Source Inverter (VSI)

A three phase 3 level voltage source inverter [15] is used. The control circuit uses an external control loop which regulates DC link voltage and an internal control loop which regulates direct axis current ($I_d$) and quadrature axis current ($I_q$), the components of the grid currents (active and reactive current). $I_d$ current reference is the output of the DC voltage external controller. Direct axis voltage ($V_d$) and quadrature axis voltage ($V_q$) outputs of the current controller are converted to three modulating signals $U_{ref\_abc}$ used by the PWM three-level pulse generator. A Phase Locked Loop (PLL) is used for synchronizing the VSI's output to the utility grid. A PLL is a circuit, which synchronizes its output with a reference or input signal in frequency as well as in phase. In synchronized state, the phase error between the system output signal and the reference signal is zero, or it remains constant. The harmonics produced by VSI are filtered by LC filter.

### B. Utility Grid

A three-phase voltage source is used as a power source for the grid. The utility grid is designed as an infinite source of power with respect to the microgrid. It can supply as well as absorb power depending on the power generation and load requirement of the microgrid. It consists of a distribution feeder and a transmission system. Two transmission lines with Pi section are incorporated to model the transmission system of the utility grid.

### C. Home Area Network (HAN)

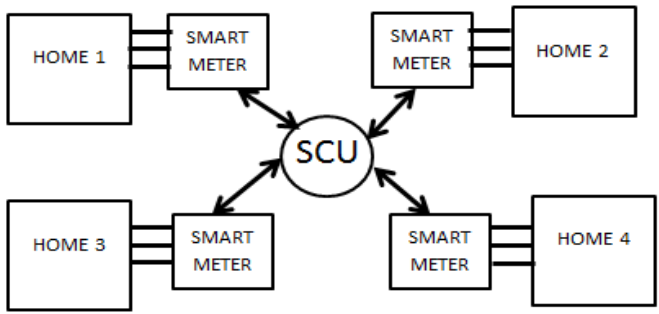

**Figure 3 Home Area Network**

The HAN, in which a communication network within the home of a residential electricity consumer allows transfer of information between homes and distribution centers, is deployed. There are four homes in the system which is supplied by three phase supply. There are three-phase and single-phase Smart Meters installed at each home for measurement of power being consumed and these smart meters are then connected to the smart control unit (SCU) via. Zigbee end device. The model of the home area network is shown in Fig.3.

### D. Smart Control Unit

There are two modes of operation of SCU depending on the mode of operation of microgrid. In islanded mode peak shaving operation is accomplished. In grid-connected mode the smart power flow is employed.

*1) Peak Shaving Operation*

SCU monitors the power demand continuously and facilitates automatic smart load shedding during the peak demand hours. The smart load shedding algorithm classifies loads as: 1) Tier-1 (must run); 2. Tier-2 (discretionary loads, i.e. can be shed for short term to reduce load peaks, or time shift them); 3. Tier-3 (emergency load shed, i.e. luxurious loads) based on historical usage data. The power consumed by the HAN is measured and communicated wirelessly. The amount of power being generated is measured at generation site using smart meters and sent to SCU. In islanded mode, during peak hours when the consumption increases and exceeds the generation, SCU disconnects the Tier-3, high power consuming luxurious loads. These loads are incorporated with a circuit breaker and relay arrangement. SCU disconnects the loads by sending the control signals to the relay circuit using wireless communication network and optimizes the load consumption resulting in peak shaving.

*2) Efficient Power Flow Management Operation*

During grid connected mode, SCU monitors the amount of power generated from the PV array and the demand of HAN in parallel. There are two conditions which govern the power flow of the microgrid:

*a) PV Generation is less than demand of HAN*

When the generated power from PV is not sufficient to full fill the load demand (even when the load is not at its peak and no power shaving is being done) then the difference between the demand and the generation of PV is supplied by grid maintaining the reliability of the supply for the HAN.

*b) PV Generation is more than load requirement*

When PV generation exceeds the demand, there is excess of power at the microgrid. Then this surplus power is fed back to the grid. Thus, the generated power is fully utilized.

## VI. SIMULATION AND RESULTS

The proposed system is modeled and simulated using MATLAB/Simulink. The characteristics of PV cell are shown in Fig.4 and Fig.5 for different irradiance levels. The output voltage of the PV array is boosted to 500V DC using a boost converter. In the VSI, the internal control loop regulates DC link voltage at 250 V and provides a voltage of 220V at the HAN end. The control circuit uses a sample time of 100 μs for voltage and current controllers as well as for the PLL [16]. Utility grid along with a 25kV/220V coupling transformer is connected to the microgrid using an isolator switch.

Fig.6 shows the Simulink model of a particular home inside the HAN, which is modeled with both single-phase and three-phase loads at each home. Fig.7 shows the Simulink model of SCU. It wirelessly fetches the measurements from the sources to calculate the available generated power and the load demand at various places in the microgrid. The SCU dispatches control signals to the circuit breakers and control units, to control power flow at home.

The wireless fetching and dispatching of measurements and control signals are simulated using from and go-to blocks in Simulink. The two operational modes of the microgrid are:

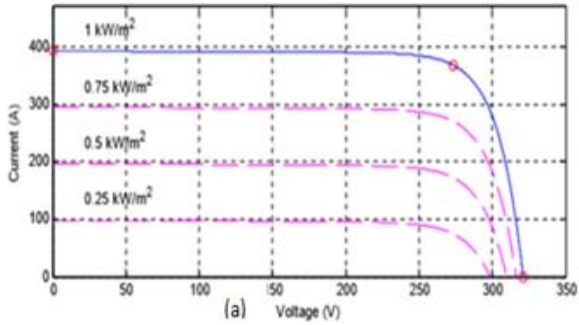
**Figure 4 I-V Characteristics of PV Cell**

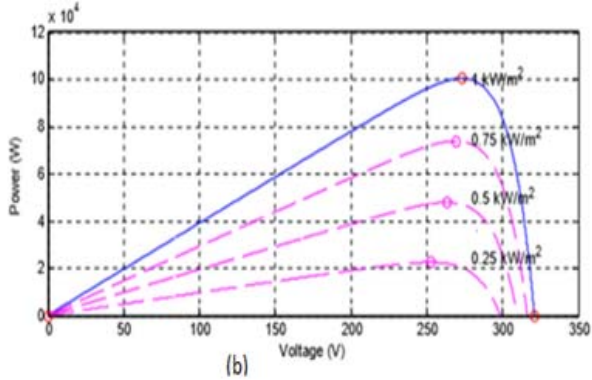
**Figure 5 P-V Characteristics of PV Cell**

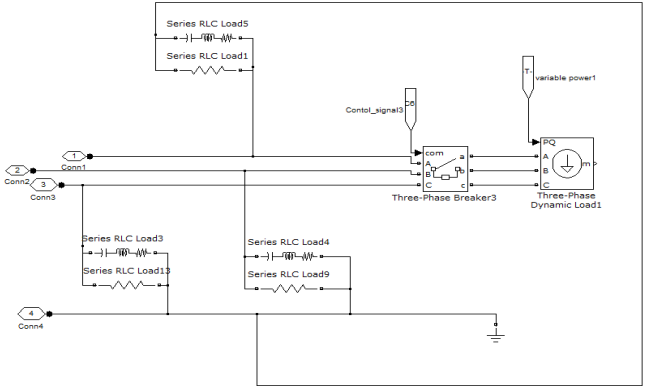
**Figure 6 Model of a home inside HAN**

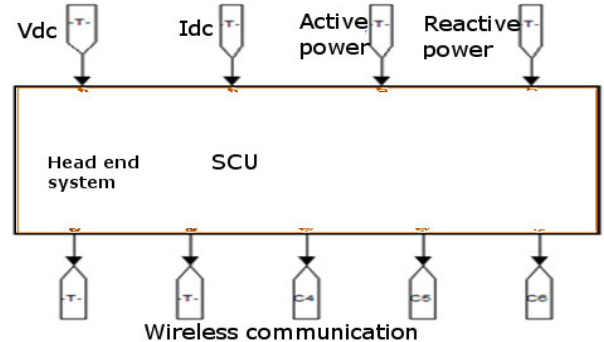
**Figure 7 SCU Block in Simulink**

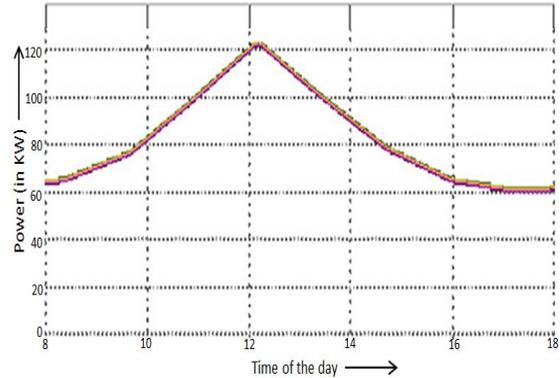
**Figure 8 Load Curve without SCU in action**

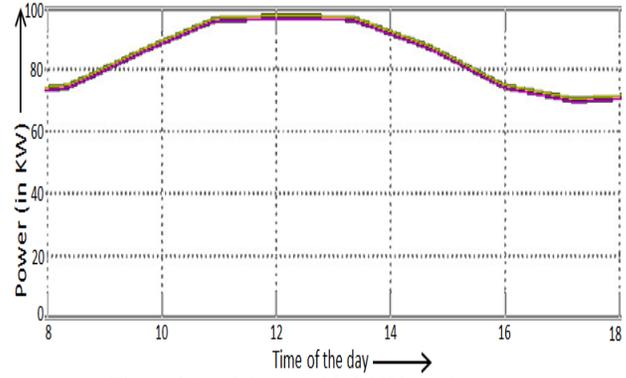
**Figure 9 Load Curve with SCU in action**

*A. Isolated Mode of Operation*

In this mode, the microgrid is disconnected from the utility grid and sustains the load demand with the power generated by the PV arrays. The overall generation capacity of the microgrid is 100 KW. The load demand curve of the microgrid on a particular day without the proposed system in action is shown in Fig. 8.

On this particular day, the peak load demand occurs between 11 am and 1pm. During these hours, the load demand exceeds the rated capacity of the microgrid, which leads to instability and eventual breakdown of the microgrid. The proposed system overcomes this problem by peak shaving, which keeps the load demand within limits (< 100 kW) even during peak demand hours as shown in Fig. 9.

*B. Grid-connected mode of operation*

In this mode, the microgrid is connected to the utility grid. The efficient power flow management operation of the SCU is evaluated under the two scenarios described in Section V. The loads are assumed to be constant in this section to distinctively show the power flow management operation under variation of the generated power.

*1) PV power generation more than the load demand*

On a bright sunny day, the irradiance is sufficiently high for the PV arrays to generate more power than the typical load demands. On such days the loads are supplied and the excess power generated is fed to the grid as shown in Fig. 10. The load demand on this particular day is around 50 KW and the generated power from PV varies from 62 KW to 85 KW. The SCU efficiently manages the power flow even when there are fluctuations in the irradiance level which can be observed during 12 hour and 15.5 hour in Fig. 10.

eliminated with the proposed way of load shedding.

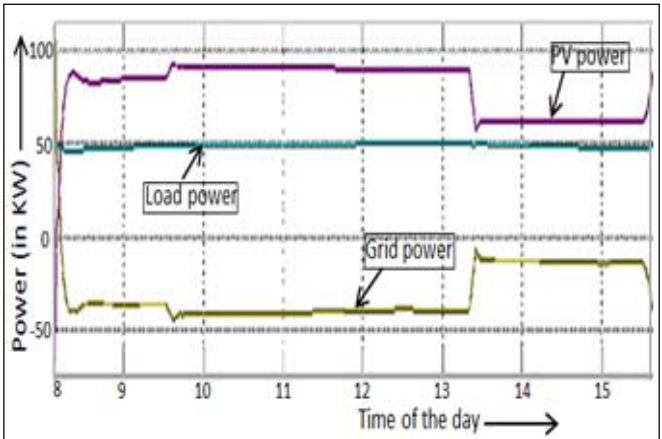

Figure 10 Power Flow Management on a Sunny Day

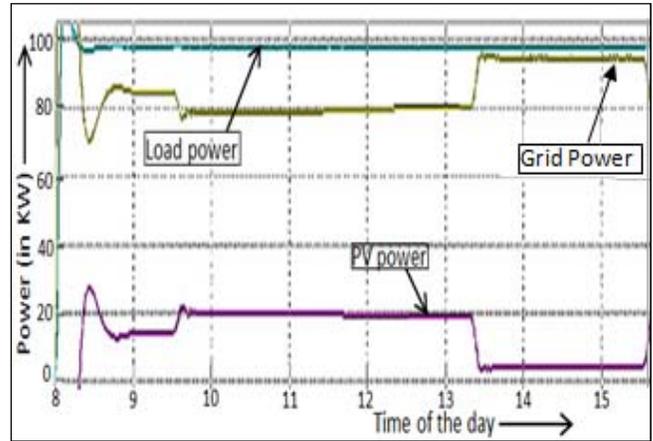

Figure 11 Power Flow Management on a Cloudy Day

*2) PV power generation more than the load demand*

Fig. 11 demonstrates the power flow management operation of the SCU on a cloudy day when the irradiance level is not high enough for the PV to generate its rated power. During such days, the SCU manages the load demand by drawing the additional required power from the grid. The load demand on this particular day is around 95 KW and the power generated from PV varies from 5 KW to 20 KW. The stability of the proposed system under very low irradiance values can be observed in Fig. 11 from 13.5 hour to 15.5 hr.

Fig. 12 shows the operation of the system for duration of 18 hours (from 6 am to 12 midnight). From morning 6 am to evening 5 pm the irradiation level is sufficient for the PV arrays to generate more than 50 KW to meet the load demand and therefore the system feeds the utility grid with the excess generated power. From 5 pm to 6 pm in the evening, the SCU partly draws from the grid to fulfill the load demand of the microgrid. The PV arrays are unable to generate power after 6 pm and therefore SCU manages to meet the load demand by fully drawing power from the utility grid.

VII. BENEFITS OF THE PROPOSED SYSTEM

The reliability of power supply in a microgrid under islanded mode is increased as with the implementation of smart load shedding (only luxurious load will be disconnected) and the rest of the loads will be supplied uninterruptedly. The requirement of energy storage for supplying peak load under islanded operation of microgrid is

The consumers get the advantage of not paying the penalty charge by decreasing the consumption in peak hours. Since the peak shaving is done automatically, the probability of error in the process decreases so the power flow in the system doesn't exceed its limits making the job of protection circuit easy and they require less maintenance. The efficient energy flow operation between the microgrid and utility grid increases the reliability of supply for HAN during grid connected operation. In case of surplus generation from renewable energy source, the power is fed back to grid. This eliminates the requirement of additional energy storage devices such as batteries, which adds additional cost to the owner. Thus, the cost of the microgrid installation gets reduced due to elimination of need of storage. Prioritizing the use of energy from the on-site renewable energy source over the utility grid (which has based load power from fossil plants), also helps in decreasing the consumption of fossil fuels, consequently advancing the clean environment goals.

VII. CONCLUSION

This paper describes a novel and smart way of load shedding which eliminates the possibility of a blackout and increases the reliability of islanded operation of a microgrid, in terms of serving the critical loads in absence of on-site stoage. An efficient energy management scheme has been presented in which the renewable source of energy; the PV array in this case is synchronized with the grid with an intelligent controller. Under grid-connected operation, the proposed Smart power flow control between PV-array and grid helps in utilizing the generated power from PV to its fullest.

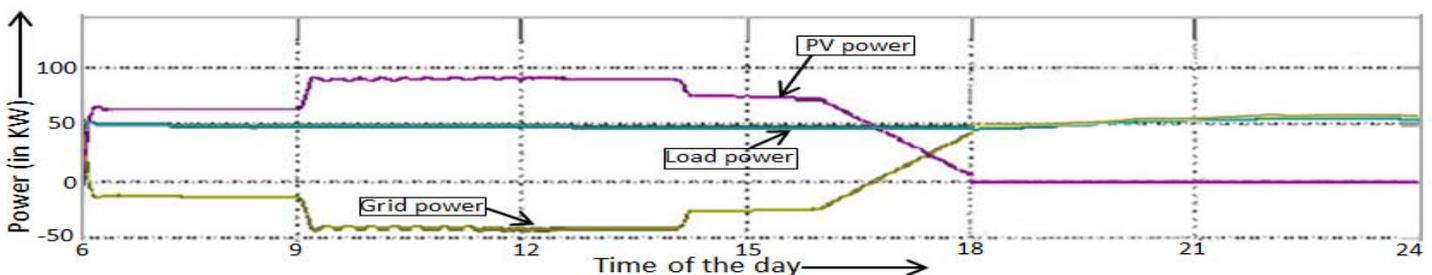

Figure 12 Power Flow Management during day and night

A future extension work can include on-site energy storage to the system to boost the reliability of the microgrid. The increased cost of the on-site storage can be counter-balanced by utilizing it for the energy arbitrage purposes.